\documentclass[11pt]{article}
\usepackage[textwidth=15.2cm,textheight=22cm]{geometry}
\usepackage{amsmath,amssymb}
\usepackage{latexsym}
\usepackage{multicol}
\usepackage{graphicx}
\usepackage{caption}
\usepackage{bm}
\allowdisplaybreaks[1]
\numberwithin{equation}{section}

\newcommand{\del}{\partial}
\newcommand{\be}{\begin{equation}}
\newcommand{\ee}{\end{equation}}
\newcommand{\ba}{\begin{eqnarray}}
\newcommand{\ea}{\end{eqnarray}}
\newcommand{\bdm}{\begin{displaymath}}
\newcommand{\nbar}[1]{\overline{#1}}
\newcommand{\edm}{\end{displaymath}}

\newcommand\fr[1]{\frac{1}{#1}}

\newcommand{\rom}[1]{\uppercase\expandafter{\romannumeral #1\relax}}
\renewcommand{\d}{\partial}
\newcommand{\E}{E_{7(7)}}
\def\parp{\partial^+}

\def\d{\partial}

\def\ba{\bar A}

\def\m#1{\mathcal#1}
\def\beq{\begin{equation}}
\def\eeq{\end{equation}}
\newcommand{\half}{\frac{1}{2}}
\newcommand{\nn}{\nonumber}

\newcommand{\ndt}{\noindent}

\newcommand{\delp}{{\partial^+}}

\newcommand{\Eei}{E_{8(8)}}

\def\bea{\begin{eqnarray}}
\def\eea{\end{eqnarray}}
\def\beas{\begin{eqnarray*}}
\def\eeas{\end{eqnarray*}}
\def\sla{\raise.15ex\hbox{$/$}\kern-.57em}

\def\parp{\partial^+}

\def\spa#1.#2{\left\langle#1\,#2\right\rangle}
\def\spb#1.#2{\left[#1\,#2\right]}

\usepackage{tikz}
\usetikzlibrary{trees}
\usepackage{siunitx}
\usepackage{varwidth}
\usepackage{pifont}
\usepackage{cancel}
\newcommand{\xdownarrow}[1]{%
  {\left\downarrow\vbox to #1{}\right.\kern-\nulldelimiterspace}
}

\usetikzlibrary{shapes.geometric, arrows, shadows}

\tikzstyle{forces} = [rectangle, rounded corners, minimum width=2cm, minimum height=0.8cm,text centered, draw=black]
\tikzstyle{spin} = [rectangle, rounded corners, minimum width=1.3cm, minimum height=0.8cm,text centered, draw=black]
\tikzstyle{theory} = [rectangle, rounded corners, minimum width=2cm, minimum height=0.8cm,text centered, draw=gray]
\tikzstyle{arrow} = [thick,->,>=stealth]
\tikzstyle{arrow1} = [thick,<->,>=stealth]
\tikzstyle{line} = [draw, -latex']

\tikzset{
  double arrow/.style args={#1 colored by #2 and #3}{ thick,
    -latex,  line width=1.1*(#1),#2, 
    postaction={draw,-latex,#3,line width=1.9*(#1/2),
                shorten <=0.7*(#1)/3,shorten >=(#1)/3}, 
  }
}
\begin{document}

\begin{titlepage}
\begin{flushright}    
{\small $\,$}
\end{flushright}
\vskip 1cm
\centerline{\Large{\bf{A hidden symmetry in quantum gravity}}}
\vskip 1.5cm
\centerline{Sudarshan Ananth$^\dagger$, Lars Brink$^*$ and Sucheta Majumdar$^\dagger$}
\vskip .7cm
\centerline{$\dagger$\,\it {Indian Institute of Science Education and Research}}
\centerline{\it {Pune 411008, India}}
\vskip 0.7cm
\centerline{$^*\,$\it {Department of  Physics, Chalmers University of Technology}}
\centerline{\it {S-41296 G\"oteborg, Sweden}}
\vskip 0.1cm
\centerline{\it {and}}
\vskip 0.1cm
\centerline{\it{Division of Physics and Applied Physics, School of Physical and Mathematical Sciences}}
\centerline{\it{Nanyang Technological University, Singapore 637371}}
\vskip 1.5cm
\centerline{\bf {Abstract}}
\vskip .5cm

\ndt The action integral contains more information than the equations of motion. We have previously shown that there are signs of an extended exceptional symmetry for $\mathcal N=8$ supergravity in four dimensions. The symmetry is such that the fields used in the Lagrangian are not representations of the symmetry. Instead one has to add representations to obtain a representation of the extended symmetry group. In this paper we discuss an extended symmetry in four-dimensional gravity which is the ``Ehlers Symmetry" in three dimensions. It cannot be spanned by the helicity states of four-dimensional gravity but it can be realised once we treat the helicity states just as field variables of the functional integral, which can be changed like variables in any integral.  We also explain how this symmetry is inherent in formulations of $\mathcal N=8$ supergravity in four dimensions through a truncation in the field space to pure gravity.  
\vfill
\end{titlepage}

\section{Introduction}

\ndt Supergravity theories show remarkable quantum properties in the sense that their perturbative expansions are finite to higher loop orders than na\"ively expected~\cite{Bern:2011qn}. Even though we expect all these theories to diverge at some loop order, it is important to understand why this is so. These phenomena must have some root in superstring theory and we expect that the study of the limiting supergravity theories will help us understand superstring theory better. When we study classical gravity in the flat limit, we look for symmetries the theory exhibits in terms of the helicity $+2$ and $-2$ fields. In the quantum case, we should study the functional integral over the action where those components are field variables that we integrate over. In the functional integral we can modify them forgetting that they are helicity fields. Hence we can ask ourselves if the functional integral has additional symmetries, over and above the spacetime symmetries that we know. 
\vskip 0.3cm
\ndt In this paper, we demonstrate signs of such a hidden symmetry in four-dimensional gravity. We use the light-cone gauge formulation in which the action is an infinite series of higher order terms and we work only up to the four-point level. Accordingly, we cannot prove that the symmetry is a symmetry of the full theory, but our experience from previous work is that if the symmetry works to this order, it is most likely to survive as a symmetry of the full theory (although we will not be able to prove this to all orders within the current formalism).
\vskip 0.3cm
\ndt The variety of additional symmetry we are interested in first appeared in the work of Cremmer and Julia~\cite{Cremmer:1979up}, who found an unexpected symmetry at the level of the equations of motion in $N=8$ supergravity. We have shown that in our approach~\cite{Brink:2008qc} this is indeed a symmetry of the full Hamiltonian and is, in some sense, on an equal footing with the maximal supersymmetry in the theory\footnote{This is because we can also use this symmetry to pin down the possible interaction terms in the Hamiltonian.}.
\vskip 0.3cm
\ndt In more recent work we have argued that the $E_{7(7)}$ symmetry should also be present in the original $d=11$ supergravity theory~\cite{Ananth:2016abv}. We then showed that the corresponding $E_{8(8)}$ symmetry, thought to be special to maximal supergravity in $d=3$, could be lifted to a symmetry of the $d=4$ theory and in principle also to the $d=11$ theory~\cite{Ananth:2017nbi}. In all these extraordinary cases we claim that the action should exhibit the symmetry. We have to carefully choose combinations of the representations used in a particular dimension to represent the symmetry but the actions do not distinguish between these.
\vskip 0.3cm
\ndt A key feature in our analysis is that the Hamiltonians in the maximally supersymmetric cases can be written as quadratic forms~\cite{Ananth:2006fh,Ananth:2015tsa,Ananth:2017xpj}. However we have also shown that for the non-supersymmetric cases, ie. pure Yang-Mills and pure gravity, this remains the case. In this paper we will investigate possible extra global symmetries in pure $d=4$ gravity. 
\vskip 0.3cm
\ndt Our light-cone formulation uses only the physical degrees of freedom. This approach is particularly well suited to the study of symmetries that are not manifest in covariant formulations~\cite{KLT}. That means that even part of the Poincar\'e symmetry is non-linearly realised. All remaining symmetries are global and the exceptional ones are described as non-linear $\sigma$-model symmetries. In the case of $E_{7(7)}$ the quotient $E_{7(7)}/SU(8)$ is non-linearly realised while the $SU(8)$ is the linear $R$-symmetry.
\vskip 0.3cm
\ndt In four-dimensional gravity no such symmetry is known but in three-dimensional gravity there is the ``Ehlers symmetry"~\cite{Ehlers}, which is an extra $SL(2,C)$ symmetry not connected to any space-time symmetry. To find the four-dimensional $\sigma$-model action with this symmetry, one must find a non-trivial change of variables in $d=4$ in the light-cone action. Here we will present an alternative method to find the $\sigma$-model, where we first study the Ehlers symmetry in $d=3$ and realize this symmetry in the four-dimensional action by means of a suitable "oxidation" procedure.
\vskip 0.3cm
\ndt
Hence we will first study $d=3$ gravity in the light-cone gauge formulation and show that there is indeed an $SU(1,1)$ $\sigma$-model symmetry. We will show that this symmetry is easily recognized only after a field redefinition. This is again a manifestation of the change of integration variables, permitted in the functional integral. 
\vskip 0.3cm
\ndt The formulation we are using is not easy to lift to $d=4$. However in a recent paper, we treated a similar problem for maximal supergravity. We found a $d=3$ formulation of the theory such that the $E_{8(8)}$ symmetry (in three dimensions) could be carefully ``oxidized" to four dimensions~\cite{Ananth:2017nbi}. We now use that analysis and truncate the superfield until it contains only the gravity degrees of freedom. Indeed, the $E_{8(8)}$ symmetry then reduces to an  $SU(1,1)$ symmetry. The formulation of pure gravity in this manner is probably one of the most impenetrable formulations of ordinary gravity and we do not recommend it for any explicit calculations but it serves its purpose, to show us the hidden symmetry.

\vskip 0.5cm

\section{$SU(1,1)$ in pure gravity in three dimensions}

\ndt In this section, we describe gravity in $d=3$ in the light-cone gauge. We do this by a straightforward dimensional reduction from $d=4$ where the light-cone formulation of gravity is well known~\cite{Scherk:1974zm,Bengtsson:1984rv,SA2}. After the reduction, we perform a suitable field redefinition that makes the Ehlers symmetry easy to write down. 

\vskip 0.3cm

\subsection{Gravity, in $d=4$, in the light-cone gauge}

\ndt With the metric $(-,+,+,+)$, we define the light-cone coordinates

\bea
x^\pm=\fr{\sqrt 2}(x^0\pm x^3)\ ;\quad x=\frac{1}{\sqrt 2}\,(\,{x_1}\,+\,i\,{x_2}\,)\ ;\quad  {\bar x} =\frac{1}{\sqrt 2}\,(\,{x_1}\,-\,i\,{x_2}\,)\ ,
\eea
with the corresponding derivatives being $\partial_\pm, \partial$ and $\bar \partial$. The Einstein-Hilbert action on a Minkowski background reads 

\bea
S_{EH}=\int\,{d^4}x\;\mathcal{L}\,=\,\frac{1}{2\,\kappa^2}\,\int\,{d^4}x\;{\sqrt {-g}}\,\,{\mathcal R}\ .
\eea

\vskip 1cm

\ndt In the light-cone gauge, the Lagrangian density in terms of the helicity states $h$ and $\bar h$ to order $\kappa^2$ reads~\cite{Ananth:2006fh} 

\bea 
\label{d=4L}
\mathcal{L}&=&\half \,\bar h\,\Box\, h\ + \, 2\,\kappa\, \bar{h}\, \delp^2\left[-\,h\,\frac{\bar{\del}^2}{\delp^2}h\,+\,\frac{\bar{\del}}{\delp}h\,\frac{\bar{\del}}{\delp}h\right] \,+\, 2\,\kappa\, h\, \delp^2\left[-\,\bar h\,\frac{{\del}^2}{\delp^2}\bar h\,+\,\frac{{\del}}{\delp}\bar h\,\frac{{\del}}{\delp}\bar h\right]\ \nn \\
&&+\ \frac{1}{{\partial^+}^2} 
    \bigg[ \partial^+ h \partial^+ \bar h\bigg]\frac{\partial \bar \partial}{{\partial^+}^2} 
    \bigg[ \partial^+ h \partial^+ \bar h\bigg]\ + \frac{1}{{\partial^+}^3} \bigg[ \partial^+ h \partial^+ \bar h\bigg] \left( \partial \bar
  \partial h \, \partial^+\bar h+ \partial^+ h \partial \bar
  \partial \bar h \right) \nonumber\\
&&
  -\frac{1}{{\partial^+}^2} 
    \bigg[ \partial^+ h \partial^+ \bar h\bigg]\, \left(2\, \partial \bar \partial h \, \bar h+ 2\, h \partial \bar
  \partial \bar h + 9\, \bar \partial h \partial \bar  h + \partial h
  \bar \partial \bar h   -  \frac{\partial \bar \partial}{\parp} h \, \partial^+\bar h
  -  \partial^+ h \frac{\partial \bar \partial}{\partial^+} \bar h 
  \right)
   \nonumber  \\
  &&
  -2 \frac{1}{\partial^+} \left[
    2 \bar \partial h\, \partial ^+\bar h 
    + h \partial^+ \bar \partial \bar h 
    - \partial^+ \bar \partial h
  \bar h \right] 
  \,h\,\partial \bar h   -2 \frac{1}{\partial^+} \left[
    2 \partial^+ h\, \partial \bar h
    + \partial^+  \partial h \, \bar h
    - h \partial^+ \partial \bar h \right] \,\bar \partial h \, \bar h\nonumber 
  \\
  &&
  -\frac{1}{\partial^+} \left[
    2 \bar \partial h\, \partial ^+\bar h 
    + h \partial^+ \bar \partial \bar h 
    - \partial^+ \bar \partial h
  \bar h \right]
  \frac{1}{\partial^+} \left[
    2 \partial^+ h\, \partial \bar h
    + \partial^+  \partial h \, \bar h
    - h \partial^+ \partial \bar h \right] \nonumber\\
  && - h\,\bar h\,\left(\partial \bar \partial h \, \bar  h + h
  \partial \bar  
  \partial \bar h + 2\, \bar \partial h \partial \bar  h 
  +3 \frac{\partial \bar \partial}{\partial^+} h \, \partial^+\bar h
  +3 \partial^+ h \frac{\partial \bar \partial}{\partial^+} \bar h 
  \right)\ .
\end{eqnarray}
\ndt The d'Alembertian in the equation above is $\Box\,=\,2\,(\,\partial\,{\bar \partial}\,-\,\partial^+\,\partial^-\,)$. The Hamiltonian to order $\kappa^2$, corresponding to the Lagrangian above can be be written in the following compact form~\cite{Ananth:2017xpj}
\bea 
\label{gqf}
\mathcal H\,=\,\int d^3 x\;\; \mathcal D\bar h\,\,\bar{\mathcal D}h\;\;,
\eea
with 
\bea \label{coderivative}
\mathcal D \bar h\,=\,\del\bar h \,+\,2\kappa\, \frac{1}{\delp^2}\,\,\big( \frac{\bar\del}{\delp} h\,\delp^3 \bar h\,-\, h\,\delp^2 \bar\del \bar h\big)\ +\ \mathcal O(\kappa^2)\ .
\eea
$\bar{\mathcal D}h$ is the complex conjugate of the expression above.

\vskip 0.5cm

\subsection{Gravity, in $d=3$, in the light-cone gauge}

\ndt We dimensionally reduce the pure gravity Lagrangian from $d=4$ to $d=3$ by setting $\partial =\bar \partial$

\bea
{\it L}&=&\fr{2}\,{\bar h}\, \Box\, h+2\kappa\,\bar h\,{\partial^+}^2\,{\biggl (}\,\frac{\partial}{\partial^+}h\,\frac{\partial}{\partial^+}h-h\,\frac{\partial^2}{{\partial^+}^2}h\,{\biggr )}\,+\,{\mbox {c.c.}}+{\it O}(\kappa^2)\ , \nn \\
&=& {\mathcal L}_0\ +\ {\mathcal L}_\kappa\ +\ {\mathcal L}_{\kappa^2}
\eea
where the d'Alembertian is now $\Box=2(\partial^2-\partial^+\partial^-)$. This expression is not suitable to search for the Ehlers symmetry. The non-linear part of it should be implemented to lowest order by $\delta h={\mbox {constant}} + \mbox{quadratic in the field}\,\ldots$, and the three-point coupling is evidently not invariant under such a transformation. In order to find the symmetry we therefore start by eliminating the cubic interaction vertices. We perform the following field redefinitions
\bea \label{redef}
h\,\rightarrow\,h'-\kappa\,{\partial^+}^2\,{\biggl (}\,\fr{\partial^+}h'\,\fr{\partial^+}h'\,{\biggr )}-2\kappa\,\fr{{\partial^+}^2}\,{\biggl (}\,{\partial^+}^3h'\,\fr{\partial^+}{\bar h}'\,{\biggr )}\ ,
\eea
with a conjugate expression for $\bar h$. These field redefinitions eliminate all cubic interaction vertices but introduce the following new quartic vertices into the Lagrangian.

\bea 
\label{extra}
&&\kappa^2 \left\{{\partial^+}^2\,{\biggl (}\,\fr{\partial^+}\bar h\,\fr{\partial^+}\bar h\,{\biggr )}+\,\fr{{\partial^+}^2}\,{\biggl (}\,{\partial^+}^3 \bar h\,\fr{\partial^+}h\,{\biggr )} \right\}\, \nn \\
&& \quad\times (\del^2\,-\, \del^+ \del^-)\, \left\{{\partial^+}^2\,{\biggl (}\,\fr{\partial^+}h\,\fr{\partial^+}h\,{\biggr )}+2\fr{{\partial^+}^2}\,{\biggl (}\,{\partial^+}^3h\,\fr{\partial^+}{\bar h}\,{\biggr )}\right\} \nn \\
&&-2\, \kappa^2 \left\{{\partial^+}^2\,{\biggl (}\,\fr{\partial^+}\bar h\,\fr{\partial^+}\bar h\,{\biggr )}+\,\fr{{\partial^+}^2}\,{\biggl (}\,{\partial^+}^3 \bar h\,\fr{\partial^+}h\,{\biggr )} \right\}\,{\partial^+}^2\,{\biggl (}\,\frac{\partial}{\partial^+}h\,\frac{\partial}{\partial^+}h-h\,\frac{\partial^2}{{\partial^+}^2}h\,{\biggr )}\, \nn \\
&& -4\, \kappa^2 \bar h\, \delp^2 \left[ \frac{\del}{\delp} \left\{ {\partial^+}^2\,{\biggl (}\,\fr{\partial^+}h\,\fr{\partial^+}h\,{\biggr )}+2\,\fr{{\partial^+}^2}\,{\biggl (}\,{\partial^+}^3h\,\fr{\partial^+}{\bar h}\,{\biggr )}\right\} \frac{\del}{\delp} h\right] \nn \\
&&+2\kappa^2\, \bar h \delp^2 \left[ \left\{{\partial^+}^2\,{\biggl (}\,\fr{\partial^+}h\,\fr{\partial^+}h\,{\biggr )}+2\,\fr{{\partial^+}^2}\,{\biggl (}\,{\partial^+}^3h\,\fr{\partial^+}{\bar h}\,{\biggr )}\right\} \frac{\del^2}{\delp^2}\, h\right] \nn \\
&& -2\, \kappa^2 \bar h \delp^2 \left[ h\, \frac{\del^2}{\delp^2} \left\{ {\partial^+}^2\,{\biggl (}\,\fr{\partial^+}h\,\fr{\partial^+}h\,{\biggr )}+2\,\fr{{\partial^+}^2}\,{\biggl (}\,{\partial^+}^3h\,\fr{\partial^+}{\bar h}\,{\biggr )}\right\} \right]\ .
\eea

\ndt The first two lines in the expression above involve $\partial^-$, which are time derivatives, and hence need to be eliminated. This is achieved by adding terms of order $\kappa^2$ to the field redefinition (\ref{redef}) which now reads
\bea 
\label{redef2}
h\,&\rightarrow&\,h'-\kappa\,{\partial^+}^2\,{\biggl (}\,\fr{\partial^+}h'\,\fr{\partial^+}h'\,{\biggr )}-2\kappa\,\fr{{\partial^+}^2}\,{\biggl (}\,{\partial^+}^3h'\,\fr{\partial^+}{\bar h}'\,{\biggr )}\  \\
&&+ \,\kappa^2 \, \frac{1}{\delp^2}\, \left\{ \frac{1}{\delp}\, \bar h' \, \delp^5 \left( \frac{1}{\delp} h'\, \frac{1}{\delp}h'\right)\right\}\ +\ 4\, \kappa^2 \, \frac{1}{\delp^2} \left\{ \delp\, \left( \delp^3 h'\, \frac{1}{\delp} \bar h'\right) \, \frac{1}{\delp} \bar h' \right\} \nn \\
&& +2\, \kappa^2 \, \frac{1}{\delp^2} \left\{ \delp^3 h'\, \frac{1}{\delp^3} \left( \delp^3 \bar h' \frac{1}{\delp} h'\right) \right\}\ +2\, \kappa^2\, \delp^2 \left\{ \frac{1}{\delp^3} \left(\delp^3 h' \, \frac{1}{\delp}\bar h' \right) \frac{1}{\delp} h' \right\}. \nn
\eea
\ndt We thus arrive at a $d=3$ Lagrangian in the following form
\bea \label{newL}
{\mathcal L'}= \mathcal L_0+\ {\mathcal L'}_{\kappa^2}\ ,
\eea

\ndt
where the new quartic interaction Lagrangian is
\bea
{\mathcal L'}_{\kappa^2} &=& {\mathcal L}_{\kappa^2}\ -\, \kappa^2 \, \delp^4\, \left(\frac{1}{\delp} \bar h \frac{1}{\delp} \bar h \right) \left[\frac{\del}{\delp}h \frac{\del}{\delp}h\,-\, h\, \frac{\del^2}{\delp^2}h \right]\ \nn \\
&&+\ 4\, \kappa^2 \delp^3 \bar h \, \frac{1}{\delp} h \left[\frac{\del}{\delp}h \frac{\del}{\delp}h\,-\, h\, \frac{\del^2}{\delp^2}h \right]\ -\ 2\, \kappa^2 \frac{1}{\delp^4}\, \left( \delp^3 h \frac{1}{\delp}\bar h\right) \, \delp^4 \bar h\, \frac{\del^2}{\delp^2}h \nn \\
&& +4 \, \kappa^2 \, \frac{1}{\delp^4}\left( \delp^3 h \frac{1}{\delp}\bar h\right) \, \delp^3 \del \bar h\, \frac{\del}{\delp} h\ -\ 2\, \kappa^2\, \frac{1}{\delp^4}\left( \delp^3 h \frac{1}{\delp}\bar h\right) \, h\, \delp^2 \del^2 \bar h \nn \\
&& -\, 8\, \kappa^2 \, \delp^2 \bar h\, \frac{\del}{\delp^2}\left( \delp^3 h \frac{1}{\delp}\bar h\right)\ \frac{\del}{\delp} h\ +\ 2\, \kappa^2 \, \delp^2 \bar h\,  \frac{1}{\delp^2}\left( \delp^3 h \frac{1}{\delp}\bar h\right)\, \frac{\del^2}{\delp^2}h \nn \\
&& +\ 4\, \kappa^2\ \delp^2 \bar h\, h\, \frac{\del^2}{\delp^4}\left( \delp^3 h \frac{1}{\delp}\bar h\right)\ ,
\eea
with the first term denoting the old quartic interaction Lagrangian. 
\vskip 0.3cm
\subsection{The $SU(1,1)$ symmetry in $d=3$}
\vskip 0.2cm
The Hamiltonian (to order $\kappa^2$) corresponding to (\ref {newL}) is
\bea
\mathcal H&=& \bar h \, \del^2 \, h\ -\ {\mathcal L'}_{\kappa^2}\ .
\eea
\ndt We can now ask if this expression could be invariant under a $\sigma$-model like symmetry of the schematic form
\be
\delta h = {\mbox {constant}} + h\\h + h\\h\\h\\h +...,
\ee
where we have not distinguished between $h$ and $\bar h$. This is quite straightforward to check and indeed we find it is invariant under the following transformations (to order $\kappa$)
\bea 
\label{del h}
\delta h&=& \frac{1}{\kappa}\, a\ - \kappa\, a\,\frac{1}{\partial^+} (\partial^+ h \bar h) -2\,\kappa\,a\,\frac{1}{{\partial^+}^2}({\partial^+}^3 h \frac{1}{\partial^+} \bar h) \nn \\
&& -\, 2\, \kappa\ \bar a\ \frac{1}{\delp^2} \left( \delp^3 h\, \frac{1}{\delp} \bar h \right)\ +\ \fr{2}\ \kappa\ \bar a\  h\ h\ ,
\eea
\vskip 0.2cm
\ndt and
\bea 
\label{del_bar h}
\delta \bar h&=& \frac{1}{\kappa}\ \bar a\ -\, \kappa\, a\, \frac{1}{\partial^+} (\partial^+ \bar h h) -2\,\kappa\,a\,\frac{1}{{\partial^+}^2}({\partial^+}^3 \bar h \frac{1}{\partial^+} h) \nn \\
&& -\ 2 \ \kappa\  a\ \frac{1}{\delp^2}\left( \delp^3 \bar h\ \frac{1}{\del^+}  h \right)\ +\ \frac{1}{2}\ \kappa\ a\ \bar h\ \bar h\ .
\eea

\ndt The commutator of two such transformations on $h$ (or $\bar h$) is
\be
[\ \delta_1\ , \delta_2\ ]\, h\ = \ 2 (\bar a_1 \ a_2\ - \bar a_2\ a_1) h\ ; \quad [\ \delta_1\ , \delta_2\ ]\, \bar h\ = \ -\,2 (\bar a_1 \ a_2\ - \bar a_2\ a_1) \bar h\ .
\ee

\ndt We can rewrite (\ref{del h}) and (\ref{del_bar h}) as two sets of transformations with parameters $a$ and $\bar a$ as follows.
\bea
L_+\, h&=& \frac{1}{\kappa}\, a\ - \kappa\, a\,\frac{1}{\partial^+} (\partial^+ h \bar h) -2\,\kappa\,a\,\frac{1}{{\partial^+}^2}({\partial^+}^3 h \frac{1}{\partial^+} \bar h), \nn \\
L_+ \, \bar h &=& -\ 2 \ \kappa\  a\ \frac{1}{\delp^2}\left( \delp^3 \bar h\ \frac{1}{\del^+}  h \right)\ +\ \frac{1}{2}\ \kappa\ a\ \bar h\ \bar h\ .
\eea 
\ndt and
\bea 
L_-\, h &=&  -\, 2\, \kappa\ \bar a\ \frac{1}{\delp^2} \left( \delp^3 h\, \frac{1}{\delp} \bar h \right)\ +\ \fr{2}\ \kappa\ \bar a\  h\ h\ , \\
L_- \, \bar h&=& \frac{1}{\kappa}\ \bar a\ -\, \kappa\, a\, \frac{1}{\partial^+} (\partial^+ \bar h h) -2\,\kappa\,a\,\frac{1}{{\partial^+}^2}({\partial^+}^3 \bar h \frac{1}{\partial^+} h) \ . \nn
\eea
\ndt We define the following $U(1)$ transformation
\be
L_0 \, h \ = \ \bar a\, a\, h\ ; \quad L_0 \bar h\ =\ - \bar a\, a\, \bar h\ .
\ee
\ndt
These transformations now satisfy an $SU(1,1)$ algebra
\bea
[\, L_+\ , \ L_-] \ =\ L_0 \,\, ;\ \quad [\, L_0\ , \ L_{\pm}]\, =\pm \, L_{\pm} .
\eea
This is the light-cone realization of the Ehlers symmetry of General Relativity. The form of the Hamiltonian used here is however not suitable to ``oxidize" to four-dimensions. This is most directly done using the Hamiltonian written as a quadratic form as in~(\ref{gqf}) and instead of trying to rewrite the Hamiltonian in such a form (which takes a lot of guesswork and partial integrations to find the final form) we will use another path.
\vskip 0.3cm
\ndt The schematic below explains how the Ehlers symmetry in three-dimensional gravity can be derived from the exceptional symmetry in maximal supergravity and subsequently lifted to four dimensions.

\vskip 1cm

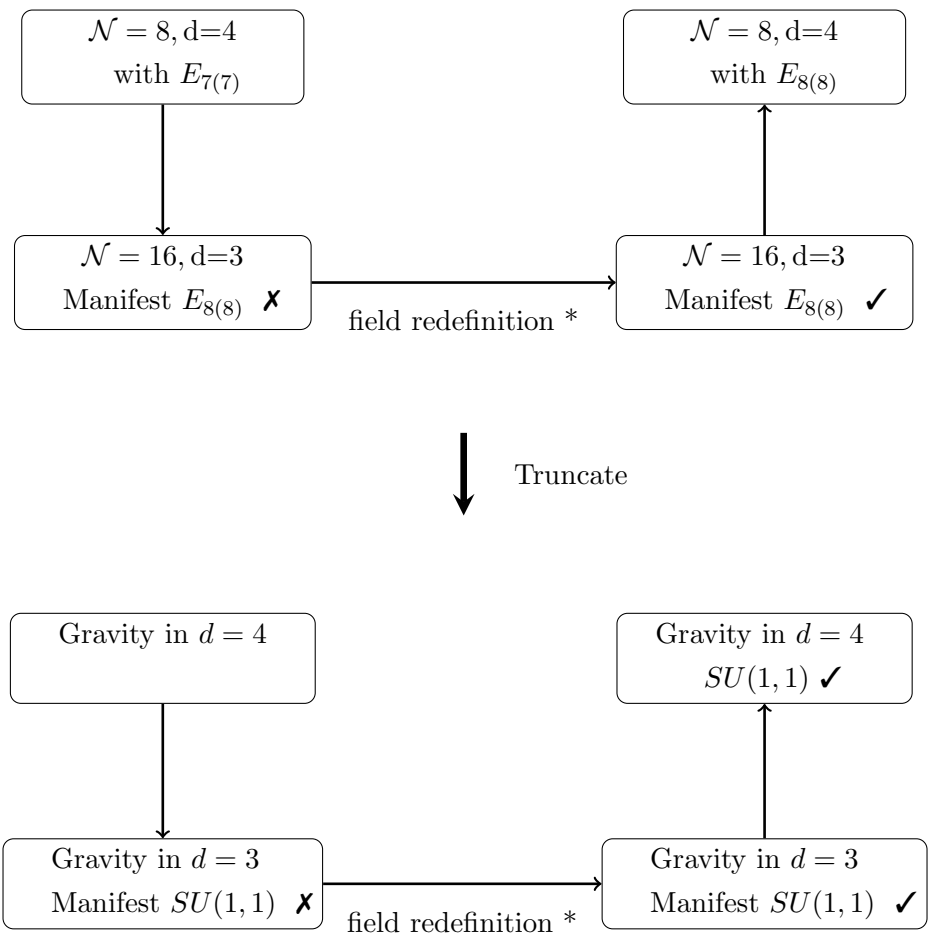
\begin{center}
\begin{figure}[h!]
\captionsetup{labelformat=empty}
\centering
\begin{tikzpicture}
\node (d4a) [forces] {\begin{varwidth}{14em} $ \qquad   \mathcal{N}=8,$\,d=4$ \qquad$   \vskip 0.1cm $\ \,  \qquad \ $with $ E_{7(7)}$  \end{varwidth}};
\node (d4b) [forces, xshift=8cm] {\begin{varwidth}{14em}$ \qquad \mathcal{N}=8,$\,d=4$ \qquad $  \vskip 0.1cm $ \ \qquad$  with $ E_{8(8)}$  \end{varwidth}} ;
\node (d3a) [forces, below of = d4a, yshift=-2.0cm] {\begin{varwidth}{14em}$\qquad \mathcal{N}=16, $\,d=3$ \qquad $  \vskip 0.1cm $\quad $ Manifest $E_{8(8)}\ \  $\ding{55} \end{varwidth}};
\node (d3b) [forces, xshift = 8cm, yshift=-3.0cm] {\begin{varwidth}{14em}$\qquad \mathcal{N}=16, $\,d=3$ \qquad $  \vskip 0.1cm $\quad $ Manifest $E_{8(8)}\ \ $\ding{51} \end{varwidth}};
\draw [->, line width=1.0pt , black] (d4a) -- (d3a);
\draw [->, line width=1.0pt, black] (d3b)-- (d4b);
\draw [->,line width=1.0pt, black] (d3a) --  (d3b) node[midway, below, yshift= -0.2cm]{field redefinition *};

\draw[->, line width=2.5pt, black, >=stealth](4,-5)--(4,-6.1)node[ midway,right, xshift= 0.5cm]{Truncate};
\node (gd4a) [forces, yshift= -8cm] {\begin{varwidth}{14em} $ \quad$ Gravity in $d=4$ $\quad$   \vskip 0.1cm $\ \,  \qquad \ $ \textcolor{white}{$SU(1,1)$ \ding{55}\qquad } \end{varwidth}};
 \node (gd4b) [forces, yshift=-8cm, xshift=8cm] {\begin{varwidth}{14em}$ \quad $Gravity in $d=4$ $\quad $  \vskip 0.1cm $ \ \qquad$  $ SU(1,1)$ \ding{51}  \end{varwidth}} ;
\node (gd3a) [forces, below of = gd4a, yshift=-2.0cm] {\begin{varwidth}{14em}$\quad $ Gravity in $d=3$$ \quad $  \vskip 0.1cm $\quad $ Manifest $SU(1,1)\ \  $\ding{55} \end{varwidth}};
\node (gd3b) [forces, xshift = 8cm, yshift=-11.0cm] {\begin{varwidth}{14em}$\quad $ Gravity in $d=3 \quad $  \vskip 0.1cm $\quad $ Manifest $SU(1,1)\ \  $\ding{51}   \end{varwidth}};
\draw [->, line width=1.0pt , black] (gd4a) -- (gd3a);
\draw [->, line width=1.0pt, black] (gd3b)-- (gd4b);
\draw [->,line width=1.0pt, black] (gd3a) --  (gd3b) node[midway, below, yshift= -0.2cm]{field redefinition *};

\end{tikzpicture}
\vskip 0.5cm
\caption{* This field redefinition eliminates cubic vertices in the action.}
\end{figure}
\end{center}

\ndt A suitable truncation of the oxidation procedure adopted in the supergravity case will help us realize the $SU(1,1)$ symmetry in $d=4$. In order to do this, we briefly present the essential points from our earlier analysis of the $\mathcal N=8$ model~\cite{Ananth:2006fh, Ananth:2017xpj}.

\section{Maximal supergravity in $d=4$}

\ndt The $\mathcal N=8$ supergravity theory in the light-cone gauge formulation is written in a $\mathcal N=8$ superspace, spanned by Grassmann variables $\theta^m$ and $\bar \theta_m$,  $m=1\,\ldots\,8$ ({\bf 8} and $\bar {\bf 8}$ of $SU(8)$), where all 256 physical degrees of freedom are captured in a single $\mathcal N=8$ superfield~\cite{BLN} 
\bea
\label{superfield}
\begin{split}
\phi\,(\,y\,)\,=&\,\frac{1}{{\parp}^2}\,h\,(y)\,+\,i\,\theta^m\,\frac{1}{{\parp}^2}\,{\bar \psi}_m\,(y)\,+\,\frac{i}{2}\,\theta^m\,\theta^n\,\frac{1}{\parp}\,{\bar A}_{mn}\,(y)\ , \\
\;&-\,\frac{1}{3!}\,\theta^m\,\theta^n\,\theta^p\,\frac{1}{\parp}\,{\bar \chi}_{mnp}\,(y)\,-\,\frac{1}{4!}\,\theta^m\,\theta^n\,\theta^p\,\theta^q\,{\bar C}_{mnpq}\,(y)\ , \\
\;&+\,\frac{i}{5!}\,\theta^m\,\theta^n\,\theta^p\,\theta^q\,\theta^r\,\epsilon_{mnpqrstu}\,\chi^{stu}\,(y)\ ,\\
\;&+\,\frac{i}{6!}\,\theta^m\,\theta^n\,\theta^p\,\theta^q\,\theta^r\,\theta^s\,\epsilon_{mnpqrstu}\,\parp\,A^{tu}\,(y)\ ,\\
\,&+\,\frac{1}{7!}\,\theta^m\,\theta^n\,\theta^p\,\theta^q\,\theta^r\,\theta^s\,\theta^t\,\epsilon_{mnpqrstu}\,\parp\,\psi^u\,(y)\ ,\\
\,&+\,\frac{4}{8!}\,\theta^m\,\theta^n\,\theta^p\,\theta^q\,\theta^r\,\theta^s\,\theta^t\,\theta^u\,\epsilon_{mnpqrstu}\,{\parp}^2\,{\bar h}\,(y)\ ,
\end{split}
\eea
\noindent with $h$ and $\bar h$ representing the graviton, ${\bar \psi}_m$ the $8$ spin-$\frac{3}{2}$ gravitinos, ${\bar A}_{mn}$ the $28$ gauge fields, ${\bar \chi}_{mnp}$  the $56$ gauginos and ${\bar C}_{mnpq}$ the $70$ real scalars. All fields are local in  
\bea
y~=~\,(\,x,\,{\bar x},\,{x^+},\,y^-_{}\equiv {x^-}-\,\frac{i}{\sqrt 2}\,{\theta_{}^m}\,{{\bar \theta}^{}_m}\,)\ .
\eea
\ndt Chiral derivatives in this space read 
\bea
d^{\,m}\,=\,-\,\frac{\partial}{\partial\,{\bar \theta}_m}\,-\,\frac{i}{\sqrt 2}\,\theta^m\,\parp\;\; ;\quad{\bar d}_n\,=\,\frac{\partial}{\partial\,\theta^n}\,+\,\frac{i}{\sqrt 2}\,{\bar \theta}_n\,\parp\  ,
\eea
\noindent and the kinematical (spectrum generating) supersymmetry generators are 
\bea
q^m_{\,+}\,=\,-\,\frac{\partial}{\partial\,{\bar \theta}_m}\,+\,\frac{i}{\sqrt 2}\,\theta^m\,\parp ;\qquad {\bar q}_{\,+\,n}=\;\;\;\frac{\partial}{\partial\,\theta^n}\,-\,\frac{i}{\sqrt 2}\,{\bar \theta}_n\,\parp\ .
\eea
\ndt To order $\kappa$, the action for ${\mathcal N}=8$ supergravity reads~\cite{BLN}

\be
\label{n=8}
-\, \frac{1}{64}\,\int\;d^4x\,\int d^8\theta\,d^8 \bar \theta\,{\cal L}\ , \nn
\ee

\bea
\label{one}
{\cal L}&=&-\bar\phi\,\frac{\Box}{\partial^{+4}}\,\phi\ + \frac{4}{3}\ \kappa \left( \frac{1}{\delp^4} \bar \phi\ {\bar \partial} \bar \del \phi\ \delp^2 \phi\ -\ \frac{1}{\delp^4} \bar \phi\ \delp  \bar \del \phi\ \delp \bar \del \phi \ +{\mbox {c.c.}}\right) .
\eea
Grassmann integration is normalized such that $\int d^8\theta\,{(\theta)}^8=1$. The Hamiltonian for the $\mathcal N=8$ theory to order $\kappa^2$ can be expressed as a Quadratic Form~\cite{Ananth:2006fh}
\bea
\label{claim}
{\cal H}~=~\frac{1}{4\,\sqrt{2}}\,(\,{\mathcal W}_{\,m}\,,\,{\mathcal W}_{\,m}\,)\ ,
\eea
where the inner product is defined as
\be
\label{inner}
(\,\phi\,,\,\xi\,)~\equiv-~2i\int d^4\!x\, d^8\theta\,d^8\,{\bar\theta}\;{\bar\phi}\,\frac{1}{{\delp}^3}\xi\ .
\ee
\ndt We note that this is unrelated to the fact that the Hamiltonian is the anticommutator of two supersymmetries. At order $\kappa$,
\bea
\label{dublew}
{\mathcal W}_m\,=-\,\frac{\partial}{\delp} {\bar q}_{+\,m}\,\phi\,-\,\kappa\,\frac{1}{\delp}\,{\Big (}\,{\bar \partial}\,{\bar d}_m\,\phi\,{{\delp}^2}\,\phi\,-\,\delp\,{\bar d}_m\,\phi\,\delp\,{\bar \partial}\,\phi\,{\Big )}\,+\,{\cal O}(\kappa^2)\ ,
\eea
\bea
\label{dublewbar}
{\nbar {\mathcal W}}^m\,=-\,\frac{\bar \partial}{\delp}\,q_+^{\,m}\,{\bar \phi}\,-\,\kappa\,\frac{1}{\delp}\,{\Big (}\,\partial\,d^m\,{\bar \phi}\,{{\delp}^2}\,{\bar \phi}\,-\,\delp\,d^m\,{\bar \phi}\,\delp\,\partial\,{\bar \phi}\,{\Big )}\,+\,{\cal O}(\kappa^2)\ .
\eea
\noindent $\mathcal W_m$ at order $\kappa^2$ is presented in~\cite{Brink:2008qc}. 
\vskip 0.3cm
\ndt The $\E/SU(8)$ transformation of the $\mathcal N=8$ supergravity theory can be written in a compact way by introducing a coherent state-like representation 

\begin{equation}
\label{E7}
\delta\phi~=~
-\frac{2}{\kappa}\,\theta^{ijkl}_{}\,\overline\Xi^{}_{ijkl}\,+\,
\frac{\kappa}{4!}\,\Xi^{ijkl}  \left(\frac{\d}{\d\eta}\right)_{ijkl}\frac{1}{\partial^{+2}}\left(e^{\eta \hat{\bar d}} \d^{+3} \phi\, e^{-\eta \hat{\bar d}}\d^{+3} \phi \right)\Bigg|_{\eta=0}\,+\, \m O(\kappa^3)\ ,
\end{equation}
where 
$$\theta^{\,a_1a_2...a_{n}}~=~ \frac{1}{n!}\theta^{a_1} \theta^{a_2 \cdots} \theta^{a_n} \ ,$$

$$
\eta\hat{\bar d} = \eta^m\frac{\bar d_m}{\d^+}~~{\rm and}~~\left(\frac{\d}{\d\eta}\right)_{ijkl} \equiv~ \frac{\d}{\d\eta^i}\frac{\d}{\d\eta^j}\frac{\d}{\d\eta^k}\frac{\d}{\d\eta^l}\ .
$$
We note that these $\E/SU(8)$ transformations do close properly to an $SU(8)$ transformation on the superfield. 
\vskip 0.5cm

\section{Truncation: from supergravity to pure gravity in $d=4$}
\vskip 0.2cm
\ndt We now note that we could set all fields, except $h$ and $\bar h$,  in the superfield to zero. The resulting expression from (\ref {claim}) is then a Hamiltonian describing pure gravity in four dimensions, in the light-cone gauge. This is another way of understanding the result in (\ref {gqf}). 
\vskip 0.2cm
\ndt In other words, the following ``superfield"
\bea
\label{superfield}
\phi\,(\,y\,)\,=\,\frac{1}{{\parp}^2}\,h\,(y)\,+\,\frac{4}{8!}\,\theta^m\,\theta^n\,\theta^p\,\theta^q\,\theta^r\,\theta^s\,\theta^t\,\theta^u\,\epsilon_{mnpqrstu}\,{\parp}^2\,{\bar h}\,(y)\ ,
\eea
\ndt furnishes us with an unnecessarily complicated description of pure gravity through the Quadratic Form defined by (\ref {claim}), (\ref  {dublew}) and (\ref {dublewbar}). 
\vskip 0.2cm
\ndt We point out that this complicated way of writing gravity was already hinted at by earlier results. In particular, we found that the light-cone Hamiltonians of both pure gravity and maximal supergravity exhibit a quadratic form structure~\cite{Ananth:2006fh, Ananth:2017xpj}. Earlier, we showed that both pure Yang-Mills theory and the maximally supersymmetric $\mathcal N=4$ Yang-Mills also exhibit this quadratic form structure~\cite{Ananth:2015tsa}. On the other hand, theories with less-than-maximal supersymmetry do not possess this property. 
\vskip 0.3cm
\ndt This form of the Hamiltonian is not suitable to look for a $\sigma$-model symmetry of the four-dimensional theory. 
When we truncate the superfield to the gravity case we see that the symmetry (\ref{E7}) disappears. In order to find a remnant of an exceptional symmetry in $d=4$ we have again to first dimensionally reduce the $N=8$ theory to three dimensions, make a field redefinition and then lift the theory back to  $d=4$ and finally perform the truncation again or make the truncation already in three dimension and then lift it. The two procedures commute.
\vskip 0.5cm

\subsection{Maximal Supergravity in three dimensions}
\vskip 0.3cm
\ndt In section $3$, we arrived at a description of $d=3$ gravity by dimensional reduction of the component Lagrangian for gravity, in the light-cone gauge. We now have a second path to the same result. 
When we dimensionally reduce the $d=4$ maximal Supergravity theory theory to $d=3$, we are left with the dependence on one transverse derivative, $ \del$. We obtain, for the action for the $d=3$ theory (up to an overall constant)

\be
\mathcal S\ =\ \int d^3 x\  d^8 \theta\  d^8 \bar \theta\ \mathcal L\ ,
\ee
\ndt 
where
\be \label{d=3 L}
\mathcal L\ =\ - \bar{\phi}\ \frac{\Box}{\delp^4}\ \phi \ + \ \frac{4}{3}\ \kappa \left( \frac{1}{\delp^4} \bar \phi\ {\partial}^2 \phi\ \delp^2 \phi\ -\ \frac{1}{\delp^4} \bar \phi\ \delp  \del \phi\ \delp  \del \phi \ +\ c.c.\right)\ ,
\ee
\vskip 0.2cm
\ndt
This theory does not show an $E_{8(8)}$ symmetry since the $SO(16)$ $R$-symmetry which is the maximal subgroup of $E_{8(8)}$ and linearly  realized does not admit vertices of odd order ($\kappa$, $\kappa^3$ etc.). It is spanned on the spinor representation $\bf 128$ for both the bosons and the fermions and there is no $\bf 1$ in the multiplication of an odd number of such spinor representations. Again we have to make field redefinitions to get rid of the three-point couplings. This was done in~\cite{Ananth:2017nbi}.
We were again led to a Hamiltonian in a quadratic form
\bea
\label{claim3}
{\cal H}^{(3)}~=~\frac{1}{4\,\sqrt{2}}\,(\,{\mathcal W^{(3)}}_{\,m}\,,\,{\mathcal W^{(3)}}_{\,m}\,)\ ,
\eea
with the superscript reminding us that we are working in $d=3$. We could now again truncate the superfield to only contain $h$ and $\bar h$ and will then recover the gravity theory in the field representation with no three-point coupling.
In~\cite{Brink:2008hv} the $E_{8,8}/SO(16)$ transformations were derived.
They read
\begin{align}\nonumber \label{E8 coset}
&\delta^{}_{E_{8(8)}/SO(16)}\,\phi~=~\frac{1}{\kappa}\,F\,+\,\kappa\,\epsilon^{m_1m_2 \dots m_8}\,\sum_{c=-2}^{2}
\left(\hat{\overline d}_{m_1m_2\cdots m_{2(c+2)}} \partial^{+c}_{}\,F\right)\\
&\quad\times
\Bigg\{ \left(\frac{\delta}{\delta\, \eta} \right)_{m_{2c+5}\cdots m_8}\,\partial^{+(c-2)} \left( e^{\eta\cdot \hat{\bar d} }  \,\partial^{+(3-c)}\phi\, e^{-\eta\cdot \hat{\bar d} } \partial^{+(3-c)}\phi\,\right)\bigg|_{\eta=0}
\,+\, \m O(\kappa^2)\Bigg\},
\end{align}
where the sum is over the $U(1)$ charges $c=2,1,0-1,-2$ of the bosonic fields, and 

\begin{eqnarray}
F&=&\,\frac{1}{{\partial^+}^2}\,\beta\,(y^-)\,\,+\,i\,\theta^{mn}_{}\,\frac{1}{\d^+}\,{\overline \beta}_{mn}\,(y^-)-\,\theta^{mnpq}_{}\,{\overline \beta}^{}_{mnpq}\,(y^-)+\nn \\
&&+\,i\widetilde\theta^{}_{~mn}\,\d^+\,\beta^{mn}\,(y^-)+\,{4}\,\widetilde\theta\,{\d^+}^2\,{\bar \beta}\,(y^-)\ ,\nn
\end{eqnarray}

$$\hat{\overline d}_{m_1m_2\cdots m_{2(c+2)}} ~\equiv~ \hat{\overline d}_{m_1}\hat{\overline d}_{m_2}\cdots\hat{\overline d}_{2(c+2)}$$
 
and

$$\widetilde\theta^{}_{~a_1a_2...a_{n}}~=~ \epsilon^{}_{a_1a_2...a_{n}b_1b_2...b_{(8-n)}}\,\theta_{}^{b_1b_2\cdots b_{(8-n)}}.$$

\vskip 0.3cm
\ndt
$F$ represents the 128 transformation parameters. We now set all the parameters in $F$, other than $\beta$ and $\bar \beta$, to zero following (\ref{superfield}).

\begin{eqnarray}
F&=&\,\frac{1}{{\partial^+}^2}\,\beta\,(y^-)\,\,+\,{4}\,\widetilde\theta\,{\d^+}^2\,{\bar \beta}\,(y^-)\ ,\nn
\end{eqnarray}
\ndt
We can then check that the exceptional $E_{8(8)}/SO(16)$ transformations (\ref{E8 coset}) break down to the $L_-$ and $L_+$ transformations in section 2, where the parameters, $a$ and $\bar a$ are identified with $\beta$ and $\bar \beta$ repectively. Similarly, the $SO(16)$ breaks down to a $U(1)$ given by $L_0$. In section $2$, we have made this realization entirely explicit. 
\vskip 0.3cm
\ndt Having established that the $d=3$ pure gravity theory possesses this symmetry, the natural next step is to ask whether we can oxidize back to four dimensions, exactly as we did with supergravity~\cite{Ananth:2017nbi}. Indeed, this can be done as explained below.

\ndt

\vskip 0.3cm
\subsection{A lift back to four dimensions}
\vskip 0.3cm
\ndt The result in (\ref {claim3}) is a particularly powerful way of realizing the Ehlers-symmetry from section $2$. This particular form of the Hamiltonian can now be oxidized back to four dimensions, while preserving this Ehlers symmetry. \ndt
This is achieved very easily by replacing all the $\partial\ ( \ =\partial_1)$ by the generalized derivative 
\be
\nabla\ \equiv\ \partial_1\ +\ i\ \partial_2\ 
\ee
\vskip 0.3cm
\ndt in the expression for $\mathcal W^{(3)}_m$ to order-$\kappa^2$~\cite{Brink:2008hv} 
\bea
&\epsilon^m\, \mathcal W^{(3)}{}_m& =\, \epsilon^m \frac{\del}{\delp}\ \bar q_m\,\phi\ \nn \\
&& +\  \frac{\kappa^2}{2} \sum^2_{c=-2}\ \frac{1}{{\delp}^{(c+4)}} \Bigg\{ \frac{\delta}{\delta a}\ \frac{\delta}{\delta b} \left( \frac{\delta}{\delta \eta}\right)_{m_1 m_2 ...m_{2(c+2)}} \bigg( E \delp^{(c+5)} \phi\  E^{-1} \bigg) \Bigg|_{a=b=\eta=0}\nn \\
&& \times \frac{\epsilon^{m_1 m_2 ...m_8}}{(4-2c)!} \left( \frac{\delta}{\delta \eta} \right)_{m_{2c+5}...m_8}\ \delp^{2c}\ \bigg( E \delp^{(4-c)} \phi E^{-1}\ \delp^{(4-c)} \phi \ \bigg) \Bigg|_{\eta=0} \Bigg\}\ ,\nn \\
&&
\eea
\ndt
where 
\be
E \equiv e^{a \hat{\del}\,+\,b\epsilon \hat{\bar q}\,+\,\eta \hat{\bar d}}\ \ \text{and}\ \ E^{-1} \equiv  e^{-a \hat{\del}\,-\,b\epsilon \hat{\bar q}\,-\,\eta \hat{\bar d}}\ , \nn
\ee
\ndt
with
\be
a\,\hat{\del}\,=\, a\frac{\del}{\delp}\ ,\quad b\,\epsilon \hat{\bar q}\,=\, b\, \epsilon^m\frac{\bar q_m}{\delp} ,\quad \eta \hat{\bar d}\,=\, \eta^m \frac{\bar d_m}{\delp} \ . \nn
\ee
\vskip 0.2cm

\ndt
The conjugate derivative, in four dimensions, enters through ${\nbar {\mathcal W}}^{(3)}$. The key point is that the $\Eei$ transformations on $\mathcal W^{(3)}$ and ${\nbar {\mathcal W}}^{(3)}$ are zero separately. This is why we could argue that also the four-dimensional action has an $\Eei$ invariance. The same argument goes through in the truncated case. This is then the statement that we find a $SU(1,1)$ internal symmetry in the $d=4$ light-cone description of the pure gravity action.

\vskip 0.5cm

\section{Conclusions}
\ndt It is well known that an action contains information beyond that in the classical equations of motion, comprised of optimal paths in field space. It is natural to ask if it also contains symmetries that are not obvious from or manifest at the level of the equations of motion~\cite{KLT}. We have shown here that such symmetries do appear in both maximally supersymmetric quantum field theories and pure gravity. The Ehlers symmetry is a well-known symmetry in three dimensional spacetime. By writing the $d=3$ Hamiltonian in a special manner we have found a way to lift that Hamiltonian to four dimensions while still exhibiting the same symmetry (which is unrelated to spacetime symmetry). In the process, we have written the Hamiltonian in several different ways seemingly getting more and more complicated but in the end finding a form that allows us to uncover this symmetry. We might in the process have found the most round-about and complex way to write the pure gravity Hamiltonian but we are not intending to use this particular form for practical calculations. The symmetries should be present even when we do not explicitly see them and hence affect calculations performed using other more convenient formalisms (this reminds us of the story of Niels Bohr and the horseshoe.)
\vskip 0.3cm
\ndt Our analysis raises the question of whether we actually know all the symmetries present in the field theories we work with. We know that Yang-Mills theory and gravity, both with and without supersymmetry, and particularly their maximally supersymmetric versions display remarkable quantum properties. We believe that we have taken a small step towards showing that there are symmetries beyond those we normally associate with these theories. We are very used, for good reasons, to working with covariant formalisms but what is the way forward when any new or hidden symmetries are only visible in non-covariant formulations or in spacetimes augmented with many extra coordinates? Our light-cone gauge formalism is democratic in the sense that all the symmetries are non-linearly implemented. This allows us, together with field redefinitions which are natural to perform in the functional integral over the action, to look for field representations which are particularly suited to these extra symmetries. We believe that there is room for further surprises.

\vskip 1cm
\ndt {\it \bf {Acknowledgments}}
\vskip 0.1cm

\ndt We thank Chris Hull for valuable discussions. The work of SA is partially supported by a DST-SERB grant (EMR/2014/000687). SM acknowledges support from a CSIR NET fellowship. LB wishes to acknowledge Aspen Physics Center where part of his work was done.

\end{document}